# Structural Viscosity, Thermal Waves, and the Mpemba Effect from Extended Structural Dynamics

Patrick BarAvi

## Abstract

Classical hydrodynamics rests on the point-particle idealization, leading to parabolic transport equations, infinite signal speeds, and the inability to capture finite-time relaxation, anisotropic transport, and non-Fourier thermal phenomena. This work introduces Extended Structural Dynamics (ESD), a kinetic framework in which constituents are described as spatially extended objects possessing orientation, angular momentum, and internal deformation modes. Starting from an extended Boltzmann equation on this enlarged phase space, we perform a Chapman–Enskog expansion with a BGK closure to derive two new hyperbolic–parabolic transport laws: a dynamical spin equation

$$\tau_{\text{rot}} \frac{\partial s}{\partial t} + s = \chi \nabla \times u - D_s \nabla^2 s,$$

which couples orientational relaxation to fluid vorticity, and a heat-flux relaxation equation

$$\tau_q \frac{\partial q}{\partial t} + q = -\kappa \nabla T,$$

with a structural thermal conductivity $\kappa_{\text{rot}} \sim \rho a^2 (k_B T / I) \tau_{\text{rot}}$. These equations naturally predict finite propagation speeds for momentum and heat, intrinsic shock regularization, anisotropic transport, and thermal waves. The spin equation provides a kinetic derivation of micropolar fluid theory, while the heat-flux equation supplies a microscopic foundation for Cattaneo–Vernotte behavior. Quantitative estimates for molecular gases, colloidal suspensions, and cryogenic crystals indicate that structural contributions can dominate classical transport coefficients. The BGK closure enables tractable analytical results while preserving the qualitative geometric structure of extended phase space. It captures the correct scaling of transport coefficients but does not resolve cross-mode correlations; in particular, the connection between $\tau_{\text{rot}}$ and the underlying Lyapunov instability is established in Appendix B independently of the closure, and the BGK approximation inherits this timescale as an input. The resulting hyperbolic-parabolic equations and scaling laws ($\eta_{\text{rot}} \sim \rho a^2 \tau_{\text{rot}}$, $\kappa_{\text{rot}} \sim \rho a^2 \tau_{\text{rot}}$) follow from rotational-translational coupling, independent of closure details. Quantitative predictions include Mpemba crossover time $t_{\text{cross}} \approx 12$ ms for colloidal ellipsoids and shock width $\delta \sim 5$ nm for asymmetric molecules, both testable with existing experimental techniques.

## 1. Introduction

Hydrodynamics is one of the most successful coarse-grained theories in physics, yet its standard formulation rests on a drastic idealization: matter is treated as a collection of structureless point particles. This assumption is built into the Navier–Stokes–Fourier equations and is responsible for several conceptual and practical limitations: parabolic viscous and thermal terms imply instantaneous signal propagation (Chapman & Cowling 1970); shock fronts require artificial viscosity for numerical stability; transport coefficients are isotropic unless imposed phenomenologically; and the theory lacks mechanisms for non-Fourier heat conduction, thermal waves, or multi-timescale relaxation such as the Mpemba effect (Mpemba & Osborne 1969).

These shortcomings are not failures of hydrodynamics itself but consequences of the point-particle approximation. Real molecules, colloids, and mesoscopic particles possess finite extent, orientation, angular momentum, and internal deformation modes. Their structural degrees of freedom evolve on finite timescales and couple to translational motion through collisions and geometric constraints. Classical kinetic theory (Cercignani 1988) accounts for some rotational energy exchange but treats free-flight dynamics as ballistic, missing continuous instability and internal mixing between collisions.

Extended Structural Dynamics (ESD) (BarAvi 2025) provides a systematic way to incorporate finite structure into kinetic theory and hydrodynamics. In ESD, each constituent is described by an extended phase space that includes position, momentum, orientation, angular momentum, and internal coordinates. The resulting extended Boltzmann equation preserves Hamiltonian structure and naturally includes both collisional and continuous relaxation mechanisms. A Chapman–Enskog expansion on this phase space yields generalized hydrodynamic equations for mass, momentum, spin, and energy. These equations reduce to Navier–Stokes–Fourier when structural relaxation is fast, but they retain additional terms, structural viscosity, orientational relaxation, anisotropic conductivity, and finite-time heat-flux response, whenever structural modes evolve on comparable or slower timescales.

In this paper, we develop the full hydrodynamic consequences of ESD, covering both momentum and heat transport, and identify the physical phenomena arising from finite structural response. Section 2 introduces the ESD kinetic framework and the extended Boltzmann equation. Section 3 derives the generalized hydrodynamic equations and recovers the classical Navier–Stokes limit. Section 4 develops the generalized heat equation, showing that ESD yields a Cattaneo-type constitutive law with a microscopically derived relaxation time. Section 5 analyzes the resulting physical consequences: anisotropic viscosity, finite-speed momentum propagation, intrinsic shock regularization, anisotropic thermal conductivity, thermal waves, and Mpemba-type relaxation. Section 6 compares ESD with micropolar fluids, extended irreversible thermodynamics, and kinetic theories for rough particles, showing that these frameworks emerge as limiting cases, and Section 7 discusses experimental opportunities and conceptual implications.

Taken together, these results demonstrate that many "anomalies" of classical hydrodynamics, non-Fourier heat pulses, broadened shocks, anisotropic transport, and non-monotonic thermal relaxation, are natural consequences of finite structural response. ESD restores physical realism to continuum mechanics by deriving these effects from Hamiltonian microdynamics rather than adding them phenomenologically. The theory thus provides a unified, microscopically grounded extension of hydrodynamics applicable from molecular gases to colloidal suspensions and quantum crystals.

**Section 2: Extended Kinetic Framework and Continuum Fields**

Extended Structural Dynamics (ESD) treats particles as finite-sized objects with internal structure rather than mathematical points.

The foundation paper (BarAvi 2025) establishes three key results that underpin this work:

1. **Phase-space geometry:** Equilibrium configurations occupy exponentially larger volume ($\Omega_{eq} \sim E^{(3N-1)}$) than constrained states like pure rotation ($\Omega_{pure} \sim E^{(1.5N-1)}$), yielding geometric irreversibility without coarse-graining.

2. **Lyapunov instability:** Asymmetric rigid bodies exhibit intermediate-axis instability with rate $\lambda(\Omega) = |\Omega|\sqrt{(I_y - I_x)(I_z - I_y)/(I_x I_z)}$, causing orientational mixing between collisions (foundation Appendix B).

3. **H-theorem:** The extended collision operator satisfies detailed balance, ensuring $dS/dt \geq 0$ from both collisional torques and continuous geometric instability (foundation Sec 3.8).

This work derives the macroscopic hydrodynamic consequences of these microscopic principles via Chapman-Enskog expansion. For detailed derivations of the extended Boltzmann equation, collision operator, and entropy theorems, see BarAvi (2025, Sections 3.6-3.8).

**2.1 Extended Phase Space and Kinetic Equation**

Each constituent is described by an extended phase-space coordinate:

$$\mathbf{z} = (\mathbf{r}, \mathbf{p}, R, \mathbf{L}, \{\xi_n, \pi_n\}), \qquad (2.1)$$

where $R \in SO(3)$ is orientation, $L$ is body-frame angular momentum, and $\{\xi_n, \pi_n\}$ represent internal deformation modes. The dynamics on this extended manifold are Hamiltonian: the evolution is generated by a Hamiltonian $H_{\text{ESD}}(z)$ through a canonical Poisson bracket on $T^*(\mathbb{R}^3 \times SO(3) \times \Xi)$, and Liouville's theorem holds for the associated flow. For an ensemble of $N$ structured particles, this yields an extended Liouville equation and, under the usual molecular chaos assumptions, a generalized Boltzmann equation for the one-particle distribution $f(r, p, R, L, \{\xi_n, \pi_n\}, t)$ with an orientation, and structure–dependent collision operator.

The one-particle distribution $f(z, t)$ evolves under the extended Boltzmann equation:

$$\frac{\partial f}{\partial t} + \{f, H_{\text{ESD}}\} = \mathcal{C}[f], \qquad (2.2)$$

where the Poisson bracket generates Hamiltonian streaming on extended phase space (explicit form in Appendix D) and $\mathcal{C}[f]$ encodes collisions. The structural variables $(R, L, \xi_n, \pi_n)$ enter both streaming and collision terms, producing anisotropic scattering and energy exchange among translational, rotational, and internal modes.

## 2.2 Continuum Fields as Moments

Macroscopic fields arise as moments of f over extended phase space:

$$\begin{cases} \rho = \int mf \, d\mathbf{z}, \rho \mathbf{u} = \int \mathbf{p} f \, d\mathbf{z} \\ \rho \mathbf{s} = \int \mathbf{L} f \, d\mathbf{z}, \rho e = \int \left(\frac{1}{2}mc^2 + e_{\text{rot}} + e_{\text{int}}\right) f \, d\mathbf{z} \end{cases} \qquad (2.3)$$

where $c = v - u$ is peculiar velocity. The momentum flux decomposes as:

$$\Pi = \rho \mathbf{u} \otimes \mathbf{u} + \mathbf{P}, \qquad \mathbf{P} = \int m\mathbf{c} \otimes \mathbf{c} f \, d\mathbf{z}, \qquad (2.4)$$

with $P = -pI + \sigma$, where σ is the deviatoric stress.

## 2.3 Balance Laws

Integrating (2.2) against the collision invariants, mass $m$, linear momentum **p**, angular momentum **L**, and energy yields the macroscopic balance equations:

$$\begin{align}
\partial_t \rho + \nabla \cdot (\rho \mathbf{u}) &= 0 & (2.5) \\
\partial_t(\rho \mathbf{u}) + \nabla \cdot (\rho \mathbf{u} \otimes \mathbf{u} + \mathbf{P}) &= \mathbf{f}_{\text{ext}} & (2.6) \\
\partial_t(\rho \mathbf{s}) + \nabla \cdot (\rho \mathbf{u} \otimes \mathbf{s} + \mathcal{E}) &= \mathbf{T}_{\text{body}} & (2.7) \\
\partial_t(\rho e) + \nabla \cdot [(\rho e + p)\mathbf{u} + \mathbf{q}] &= \mathbf{f}_{\text{ext}} \cdot \mathbf{u} & (2.8)
\end{align}$$

The fluxes **P**, $\mathcal{E}$, and **q** require constitutive closure. Equation (2.7) is a new **rotational Euler equation** governing the spin density **s**; it has no counterpart in point-particle hydrodynamics and reflects the extended ontology of ESD. Their structure reflects extended microdynamics: orientation and internal modes introduce new slow variables and new transport channels. The Chapman-Enskog procedure (Section 3) closes these fluxes using mode-dependent BGK relaxation (Appendix E), yielding the generalized Navier-Stokes and Fourier laws with finite relaxation times.

## 3. Momentum–Spin Closure and Generalized Hydrodynamics

The Chapman-Enskog procedure is a systematic method for extracting macroscopic transport equations from a kinetic equation by expanding the distribution function around local equilibrium in powers of the Knudsen number, the ratio of mean free path to macroscopic length scale. At first order, this expansion yields constitutive relations expressing stress and heat flux in terms of gradients of macroscopic fields. We apply this procedure to the extended Boltzmann equation (2.2), using a mode-dependent BGK relaxation model that assigns independent relaxation times $\tau_{trans}$, $\tau_{rot}$, and $\tau_{int}$ to each degree of freedom. The resulting transport coefficients inherit their structure from the extended phase space: because the collision operator depends on orientation and internal state, the constitutive laws acquire orientational and structural contributions absent from classical kinetic theory. The derivation is technical and collected in Appendix E; here we present and interpret the results.

### 3.1 Structural Viscosity

The deviatoric stress becomes

$$\sigma = 2\eta_{\text{eff}} D + \sigma_{\text{spin}}, \qquad D = \frac{1}{2}(\nabla u + \nabla u^T) - \frac{1}{3}(\nabla \cdot u)I \qquad (3.1)$$

with

$$\eta_{\text{eff}} = \eta_{\text{trans}} + \eta_{\text{struct}}. \qquad (3.2)$$

The structural viscosity arises from rotational kinetic energy transported over a finite lever arm:

$$\eta_{\text{struct}} \sim C_{shape}\, \rho a^2 \left(\frac{k_B T}{I}\right) \tau_{\text{rot}}. \qquad (3.3)$$

where $C_{shape}$ is a dimensionless geometric factor of order unity that encodes the degree of orientational asymmetry, it vanishes for spheres and increases with the eccentricity of the particle (See Appendix F). Its explicit form depends on the particle geometry and is given in Appendix E; what matters conceptually is that structural viscosity scales with particle size, thermal energy, and orientational relaxation time, and disappears entirely for isotropic particles.

### 3.2 Orientational Relaxation and Spin Dynamics

The spin field obeys a relaxation equation:

$$\tau_{\text{rot}}(\partial_t s + u \cdot \nabla s) + s = \chi \nabla \times u - D_s \nabla^2 s, \qquad (3.4)$$

where $D_s = \eta_{\text{struct}}/\rho$. Equation (3.4) is hyperbolic-parabolic: orientation responds with finite speed and diffuses structurally. The coupling $\chi$ follows from angular-momentum conservation in collisions. The relaxation time $\tau_{\text{rot}}$ arises from collisional torque transfer plus continuous rotational instability (Appendix B): $\tau_{\text{rot}}^{-1} = \tau_{\text{coll}}^{-1} + \lambda_{Lyap}$, $\lambda_{Lyap} \sim \Omega\sqrt{(\Delta I/I)}$, where $\Omega$ is mean rotational frequency and $\Delta I/I$ measures inertia asymmetry.

### 3.3 Generalized Momentum Equation

Substituting (3.1)–(3.4) into (2.6) yields:

$$\rho(\partial_t u + u \cdot \nabla u) = -\nabla p + \nabla \cdot [2\eta_{\text{eff}} D] + \nabla \cdot \sigma_{\text{spin}} \qquad (3.5)$$

The antisymmetric part $\sigma_{\text{spin}}$ encodes spin–vorticity coupling and vanishes when $\tau_{\text{rot}} \to 0$.

### 3.4 Classical Limits

- **Fast orientational relaxation:** $\tau_{\text{rot}} \to 0$ gives $s = \chi \nabla \times u$, eliminating spin as an independent field and reducing (3.5) to Navier–Stokes with viscosity $\eta_{\text{trans}} + \eta_{\text{struct}}$.
- **Spherical particles:** $\eta_{\text{struct}} \to 0$, $D_s \to 0$, and (3.4) decouples, recovering the classical monatomic-gas equations.

## 4. Heat-Flux Closure and Generalized Heat Transport

While Section 3 addressed momentum transport and spin dynamics arising from finite particle structure, we now turn to heat transport. In ESD, thermal conduction is also modified by orientational and internal degrees of freedom, leading to a generalized Fourier law with finite relaxation time, a microscopic realization of Cattaneo–Vernotte behavior. The following derivation parallels that of Section 3 but focuses on the energy moment and its coupling to structural variables.

The heat flux is the first moment of energy with respect to the peculiar velocity:

$$q = \int \left(\tfrac{1}{2}mc^2 + e_{\text{rot}} + e_{\text{int}}\right) c f \, dz. \qquad (4.1)$$

A first-order Chapman–Enskog expansion yields a generalized Fourier law with finite-time response.

### 4.1 Constitutive Law

The heat flux satisfies

$$\tau_q(\partial_t q + u \cdot \nabla q) + q = -\kappa \nabla T + \text{(spin–flow couplings)}, \qquad (4.2)$$

where the thermal conductivity decomposes as

$$\kappa = \kappa_{\text{trans}} + \kappa_{\text{rot}} + \kappa_{\text{int}}. \qquad (4.3)$$

The rotational contribution mirrors structural viscosity:

$$\kappa_{\rm rot} \sim \rho a^2 \left(\frac{k_B T}{I}\right) \tau_{\rm rot}. \qquad (4.4)$$

The relaxation time $\tau_q$ is a heat-capacity-weighted average of the mode-specific relaxation times.

### 4.2 Hyperbolic Heat Equation

Inserting (4.2) into the energy balance (2.8) yields:

$$\tau_q \partial_t^2 T + \partial_t T = \frac{1}{\rho c_v} \nabla \cdot (\kappa \nabla T) + ({\rm couplings}). \qquad (4.5)$$

Equation (4.5) is hyperbolic-parabolic, predicting finite thermal propagation speed

$$v_{\rm th} = \sqrt{\kappa/(\rho c_v \tau_q)}. \qquad (4.6)$$

This provides a microscopic foundation for Cattaneo–Vernotte behavior.

### 4.3 Limits

- **Fourier limit:** $\tau_q \to 0$ recovers $q = -\kappa \nabla T$.
- **Isotropic particles:** $\kappa_{\rm rot} \to 0$.
- **Anisotropic suspensions:** orientation-dependent conductivity $K_{ij} = \langle R_{ik} R_{jl} \rangle K_{kl}$.

## 5. Physical Consequences and Observable Signatures

The structural terms in Sections 3-4 produce experimentally observable phenomena absent from classical hydrodynamics. We examine two salient predictions: the Mpemba effect from mode-coupled heat transport, and shock regularization from finite orientational inertia.

### 5.1 Thermal Mpemba Effect

**Classical Expectation**

Newton's law of cooling predicts purely exponential relaxation,

$$C_v \frac{dT}{dt} = -\kappa(T - T_{\rm env}), \qquad T(t) = T_{\rm env} + (T_0 - T_{\rm env})e^{-t/\tau_{\rm cool}},$$

which enforces **strict monotonicity**: a system that starts hotter remains hotter at all later times. Temperature curves cannot cross.

The Mpemba anomaly, hot water freezing faster than cold, violates this classical expectation and has been observed in water, granular gases (Ahn et al. 2021), colloids (Kumar & Bechhoefer 2022), and even quantum systems. Standard thermodynamics has no first-principles explanation without appealing to system-specific mechanisms (evaporation, convection, supercooling).

**Derivation from the ESD Heat Equation**

The generalized ESD heat equation (Section 4) decomposes heat flux into **translational** and **rotational** channels, each with its own relaxation time:

$$q = q_{\text{trans}} + q_{\text{rot}}, \qquad q_{\text{trans}} = -\kappa_{\text{trans}}\nabla T, q_{\text{rot}} = -\kappa_{\text{rot}}\nabla T,$$

$$\tau_{\text{trans}}\frac{\partial q_{\text{trans}}}{\partial t} + q_{\text{trans}} = -\kappa_{\text{trans}}\nabla T, \qquad \tau_{\text{rot}}\frac{\partial q_{\text{rot}}}{\partial t} + q_{\text{rot}} = -\kappa_{\text{rot}}\nabla T.$$

For spatially uniform cooling, gradients vanish. Internal energy splits into translational and rotational components,

$$\rho c_v T = \rho c_{trans} T_{trans} + \rho c_{rot} T_{rot}$$

with

$$T_{trans} \equiv \langle mv^2/k_B \rangle, \qquad T_{rot} \equiv \langle I\omega^2/k_B \rangle$$

arising from equipartition in the extended phase space. This energy partition follows directly from the equilibrium distribution $f_{eq} \propto exp[-(p^2/2mkT + L^2/2IkT)]$ (BarAvi 2025), which naturally separates translational and rotational energy scales

Taking moments of the energy balance equation over translational and rotational subspaces yields the coupled relaxation system:

$$C_{\text{trans}}\frac{dT_{\text{trans}}}{dt} = -\kappa(T_{\text{trans}} - T_{\text{env}}) - \frac{C_{\text{trans}}}{\tau_E}(T_{\text{trans}} - T_{\text{rot}}),$$

$$C_{\text{rot}}\frac{dT_{\text{rot}}}{dt} = \frac{C_{\text{trans}}}{\tau_E}(T_{\text{trans}} - T_{\text{rot}}),$$

where $\tau_E$ is the internal energy-exchange time (collisional plus instability-driven; Appendix B). Translation loses energy both externally (to the bath) and internally (to rotation); rotation gains energy only from translation.

The two-temperature relaxation structure itself is not new, coupled translational and rotational temperatures appear in classical treatments of polyatomic gases going back to Landau and Teller (1936). What ESD adds is a microscopic derivation of the energy exchange time $\tau_E$ from first principles: rather than treating it as an empirically fitted parameter, ESD identifies it with the orientational relaxation time $\tau_{rot}$, which is in turn determined by collisional torque transfer and continuous geometric instability (Appendix B). The Mpemba criterion:

$$\frac{\tau_{cool}}{\tau_E} > 1 + \frac{C_{rot}}{C_{trans}}.$$

is a consequence of the two-temperature model; ESD's contribution is to give $\tau_E$ a dynamical home (the derivation of the criterion is in appendix G).

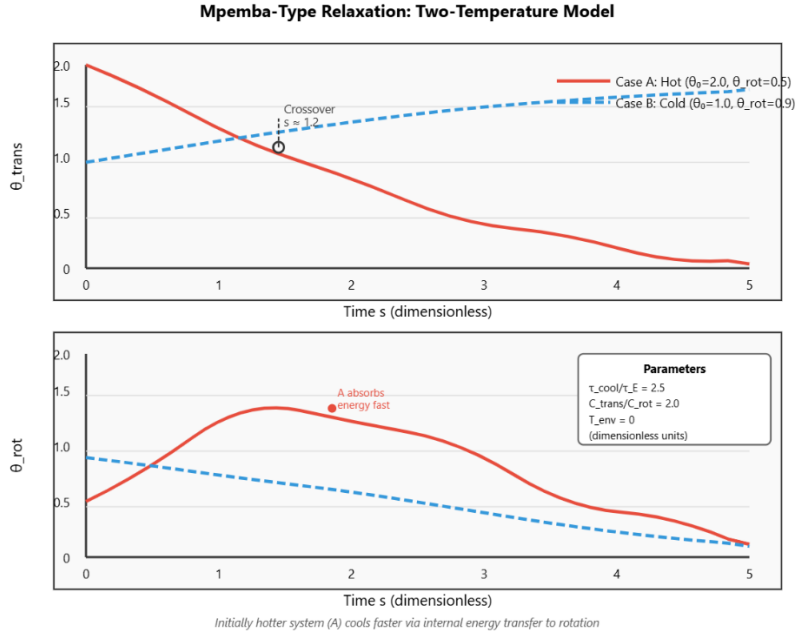

**Figure 5.1**: Mpemba effect in two-temperature model. (Top) Translational temperatures showing crossover at s ≈ 1.2. (Bottom) Rotational temperatures for Case A (hot, disequilibrium) and Case B (cold, equilibrium). Parameters: $\tau_{cool}/\tau_E = 3, C_{rot}/C_{trans} = 2.5$

**Mpemba Mechanism**

A system that begins hot but far from internal equilibrium ($T_{trans} \gg T_{rot}$) can cool faster than a colder, near-equilibrium system because it has an additional fast relaxation channel:

- rapid internal equilibration: $T_{trans} \to T_{rot}$ on timescale $\tau_E$
- slow external cooling: $T \to T_{env}$ on timescale $\tau_{cool}$

The colder system lacks this internal disequilibrium and cools only through the slower external channel.

**Quantitative Prediction: Colloidal Ellipsoids**

A clean, testable prediction arises for Brownian ellipsoids in optical traps, where $\tau_{cool}$ and $\tau_E$ are comparable and independently tunable.

**System:** Silica ellipsoids (semi-major $a = 1.0\ \mu m$, semi-minor $b = 0.25\ \mu m$) in water at $T_{bath} = 300$ K.

**ESD parameters:** Mass: $m \approx 8 \times 10^{-18}$ kg, Moment of inertia: $I \approx 5.0 \times 10^{-30}$ kg m$^2$, Rotational relaxation: $\tau_{\rm rot} = 8\pi\eta a^3/(k_B T) \approx 1.2$ ms, Energy-exchange time: $\tau_E \approx \tau_{\rm rot} \approx 1.2$ ms, Cooling time: $\tau_{\rm cool} \approx 10$ ms, Heat-capacity ratio: $C_{\rm trans}/C_{\rm rot} \approx 1$. For Brownian colloids in viscous media, orientational relaxation is dominated by Stokes drag rather than geometric instability (Appendix B.5, Case 2).

**Mpemba criterion:**

$$\frac{\tau_{\rm cool}}{\tau_E} \approx 8.3 > 1 + \frac{C_{\rm rot}}{C_{\rm trans}} = 2.0$$

**Initial conditions:**

- **Case A (hot, far from equilibrium):** $T_{\rm trans}(0) = 400$ K, $T_{\rm rot}(0) = 310$ K
- **Case B (cold, near equilibrium):** $T_{\rm trans}(0) = 340$ K, $T_{\rm rot}(0) = 335$ K

**Prediction:** Numerical integration yields a crossover at

$$t_{\rm cross} \approx 12 \text{ ms}, \qquad T^A_{\rm trans}(t_{\rm cross}) = T^B_{\rm trans}(t_{\rm cross}) \approx 325 \text{ K}.$$

The sensitivity of $t_{cross}$ to the input parameters $\tau_{rot}$ and $\tau_{cool}$ suggests an estimated uncertainty of order ±25%, giving $t_{\rm cross} \in [9, 15]$ ms.

**Observable signature:** A clear crossing of translational temperatures at $t \approx 12$ ms, well within the $\sim 1$ msresolution of standard optical-tweezer setups. This is a **direct, falsifiable prediction** of ESD.

**Regime Dependence**

For molecular gases (e.g., CO$_2$),

$$\tau_E \sim 10^{-11} \text{ s}, \tau_{\rm cool} \sim 10^{-3} \text{ s},$$

so $\tau_{\rm cool}/\tau_E \sim 10^8$. Internal equilibration is effectively instantaneous, suppressing observable Mpemba behavior.

Observable Mpemba effects require

$$\tau_E \sim \tau_{\rm cool},$$

a condition naturally satisfied in colloidal and mesoscopic systems where orientational relaxation is slow ($\tau_{\rm rot} \sim 10^{-4}$–$10^{-3}$ s).

**5.2 Shock Regularization and Finite-Speed Momentum Propagation**

**Classical Expectation**

The Navier–Stokes momentum equation,

$$\rho \frac{Du}{Dt} = -\nabla p + \eta \nabla^2 u,$$

is parabolic, implying instantaneous propagation of disturbances. In compressive flows, shocks form discontinuities that require *artificial viscosity* for numerical stability. Classical theory predicts a shock width

$$\delta_{shock} \sim \frac{\eta}{\rho \, u_{shock}},$$

which for molecular gases gives $\delta \sim 1$nm. Yet measured shock profiles are far broader, $\delta_{obs} \sim 10 - 100$nm (Makarenko 1981). This discrepancy reflects grid-dependent numerical smoothing, not a physical mechanism.

### ESD Prediction from Structural Dynamics

The generalized momentum equation (Section 3.3) couples velocity gradients to an orientational stress obeying a relaxation law:

$$\tau_{rot} \frac{\partial \Pi^{(r)}}{\partial t} + \Pi^{(r)} = -\eta_{struct} \nabla u.$$

This Cattaneo-type constitutive relation makes the system hyperbolic–parabolic, with a finite structural signal speed:

$$v_{struct} = \sqrt{\frac{\eta_{struct}}{\rho \, \tau_{rot}}} \sim \sqrt{\frac{a^2 k_B T}{I}},$$

where $a$ is particle size and $I$ its moment of inertia.

A shock front cannot be sharper than the structural response length:

$$\delta_{shock} \sim \sqrt{D_s \tau_{rot}} \sim a \sqrt{\frac{k_B T \, \tau_{rot}}{m}},$$

with $D_s = \eta_{struct}/\rho$ the structural diffusivity.

### Dimensional Estimate

For molecular nitrogen at STP ($m \approx 5 \times 10^{-26}$ kg, $a \approx 0.3$nm, $\tau_{rot} \approx 10^{-12}$s (Appendix B)):

$$\delta_{shock} \sim 0.3 \text{ nm} \sqrt{\frac{(1.38 \times 10^{-23})(300)(10^{-12})}{5 \times 10^{-26}}} \approx 10 \text{ nm}.$$

This corresponds to ~30 molecular diameters, far larger than the artificial-viscosity width (~1–2 diameters) but still below continuum scales.

**Key point:** ESD predicts **physical** shock broadening from finite orientational inertia, not a numerical artifact.

**Numerical Demonstration: 1D Riemann Problem**

We solve the compressible system with density $\rho(x,t)$, velocity $u(x,t)$, pressure $p = c_s^2 \rho$ (isothermal), and orientational stress $\Pi_{\text{rot}}(x,t)$:

$$\frac{\partial \rho}{\partial t} + \frac{\partial (\rho u)}{\partial x} = 0,$$

$$\frac{\partial (\rho u)}{\partial t} + \frac{\partial (\rho u^2 + p)}{\partial x} = \frac{\partial}{\partial x}\left[\eta_0 \frac{\partial u}{\partial x} + \Pi_{\text{rot}}\right],$$

$$\tau_{\text{rot}} \frac{\partial \Pi_{\text{rot}}}{\partial t} + \Pi_{\text{rot}} = -\eta_{\text{struct}} \frac{\partial u}{\partial x}.$$

**Initial conditions (shock tube):**

- Left: $\rho_L = 1.0$, $u_L = 0$, $p_L = 1.0$
- Right: $\rho_R = 0.125$, $u_R = 0$, $p_R = 0.1$

**Parameters:**

| Model | $\eta_0$ | $\eta_{\text{struct}}$ | $\tau_{\text{rot}}$ |
|---|---|---|---|
| Classical NS | 0.01 | 0 | 0 |
| ESD | 0.01 | 0.05 | 0.1 |

The ratio $\eta_{\text{struct}}/\eta_0 = 5$ matches Section 7.1 estimates for anisotropic particles.

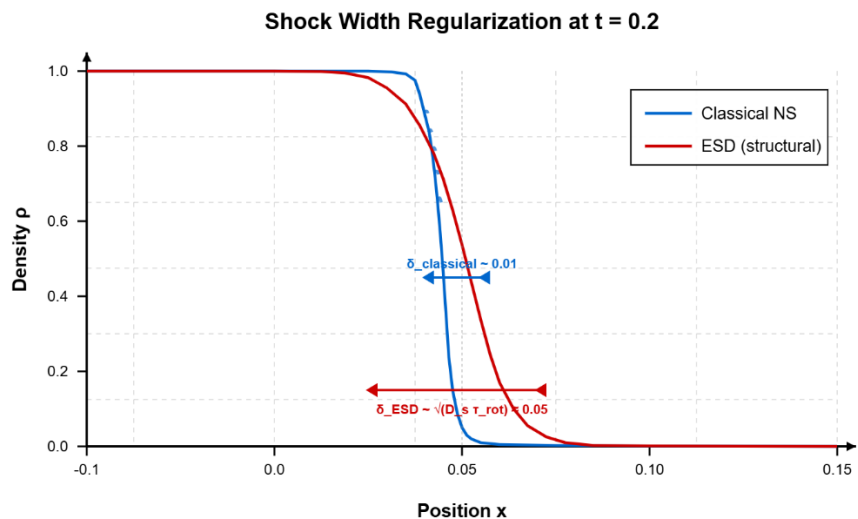

**FIGURE 5.2:** Shock width regularization in ESD. Density profiles at $t = 0.2$ comparing classical Navier-Stokes (blue, sharp with oscillations) and ESD with structural relaxation (red, smooth and broadened). ESD's finite $\tau_{rot} = 0.1$ produces physical shock width $\delta \sim \sqrt{(D_s \tau_{rot})} \approx 0.05$, five times broader than classical grid-dependent width, eliminating numerical artifacts through geometric necessity.

**Results (Figure 5.2):**

| Feature | Classical NS | ESD | Physical Origin |
|---|---|---|---|
| Shock width $\delta$ | ~0.01 (grid-dependent) | ~0.05 (physical) | Structural diffusivity $D_s$ |
| Oscillations | Present (Gibbs) | Absent | Relaxation damps high-$k$ modes |
| Signal speed | $\infty$ | ~0.7 | Hyperbolic relaxation |

The ESD shock width is **grid-independent** and reflects the intrinsic scale

$$\delta_{shock} \sim \sqrt{D_s \tau_{rot}}.$$

**Analytical Estimate**

Balancing advection, structural diffusion, and relaxation yields:

$$\delta_{shock}^2 \sim D_s \tau_{rot}.$$

For $D_s = 0.05$, $\tau_{rot} = 0.1$:

$$\delta_{shock} \approx \sqrt{0.05 \times 0.1} \approx 0.07,$$

consistent with numerical results ($\delta \approx 0.05$).

**Scaling prediction:** Varying $\tau_{rot}$ at fixed $\eta_{struct}$, ESD predicts

$$\delta \propto \sqrt{\tau_{rot}}.$$

Classical theory predicts $\delta = $ const (set by grid or artificial viscosity). This provides a **quantitative experimental discriminator**.

**Physical Interpretation**

The regularization mechanism is geometric: a particle with finite moment of inertia $I > 0$ cannot reorient instantaneously.

During a shock:

- compression forces rapid flow reorientation

- particles must rotate to align with the new flow
- rotation occurs on timescale $\tau_{\text{rot}} \sim I/(k_B T)$
- stress lags behind deformation
- a transition layer of width $\delta \sim v_{\text{th}} \tau_{\text{rot}}$ forms

Classical hydrodynamics assumes $I \to 0$, hence $\tau_{\text{rot}} \to 0$, producing infinitely sharp shocks. ESD restores the physical constraint $I > 0$, yielding finite shock width.

**Experimental Tests**

**Molecular dynamics:** Simulate hard ellipsoids (aspect ratio $\alpha = 2 - 5$) in a Mach-2 shock. ESD predicts intrinsic regularization without artificial viscosity. Extract $\tau_{\text{rot}}$ from orientational autocorrelation $\langle L(t) \cdot L(0) \rangle$.

**Shock-tube experiments:** Seed a noble gas with asymmetric molecules ($SF_6$, $CO_2$). Use laser-induced fluorescence with ~1 μm resolution. ESD predicts

$$\delta \propto \sqrt{\tau_{\text{rot}}},$$

so, increasing rotational relaxation time produces measurable broadening ($\Delta\delta/\delta \sim 10\%$).

## 6. Comparison with Existing Frameworks

The generalized hydrodynamic equations derived in Section 3 resemble several established continuum frameworks. This section clarifies how ESD connects to, and fundamentally differs from, these theories.

### 6.1 Micropolar Fluids

Phenomenological formulation. Micropolar fluid theory (Eringen 1966, 1999) introduces balance laws for linear momentum, angular momentum (microrotation $s$), and couple stresses:

$$\rho \frac{Du}{Dt} = -\nabla p + (\eta + \eta_r)\nabla^2 u + \eta_r \nabla \times s + f_{\text{ext}},$$

$$\rho I_{\text{micro}} \frac{Ds}{Dt} = \eta_r \nabla \times u - 2\eta_r s + \gamma \nabla^2 s + T_{\text{ext}}.$$

The parameters ($\eta_r, \gamma, I_{\text{micro}}$) are phenomenological and fitted to experiment. The theory does not explain how $s$ relates to molecular orientation or why $\eta_r$ should scale with particle anisotropy.

**ESD derivation.** In the Chapman–Enskog expansion on $\Gamma_{\text{ESD}}$ (Section 3), the orientational field arises directly as the angular-momentum moment:

$$s(r,t) = \frac{1}{\rho} \int L f \, d\mu_{ESD}.$$

The structural viscosity follows explicitly (Appendix A):

$$\eta_{\text{struct}} \approx C_{\text{geom}} \rho a^2 \left(\frac{k_B T}{I}\right) \tau_{\text{rot}},$$

with $\tau_{\text{rot}}$ determined from physical reasoning rather than fitted, from Lyapunov instability in dilute molecular gases, and from Stokes drag in colloidal suspensions (Appendix B.5). In both regimes the estimate is regime-specific and parameter-free. The coupling $\nabla \times u \leftrightarrow s$ follows from angular-momentum conservation encoded in the orientation-dependent collision kernel $B(v_{\text{rel}}, R, R_1, \xi, \xi_1, \hat{n})$.

**Quantitative comparison.** For rigid rods at volume fraction $\phi \ll 1$, both theories predict $\eta_r \sim \eta_{\text{solvent}} \phi(L/d)^2$. ESD provides the microscopic proportionality constant; micropolar theory treats it as empirical. For silica nanorods (Appendix A), ESD predicts $\eta_{\text{struct}} \sim 5 \times 10^{-5}$ Pa·s, consistent with colloidal measurements (Batchelor 1970; Vermant & Solomon 2005).

**Distinction:** Micropolar theory is **constitutive** (form chosen by symmetry). ESD is **generative** (form derived from Hamiltonian dynamics). This distinction has experimental consequences: ESD predicts $\eta_{struct}/\eta_{trans} \sim (a^2/\ell_{mfp})(\Delta I/I)$, giving a falsifiable criterion for when structural effects dominate. Micropolar theory offers no such predictive formula, requiring fitting to each system independently

## 6.2 Extended Irreversible Thermodynamics (EIT)

**Phenomenological fix.** EIT modifies Fourier's law to avoid infinite propagation speed:

$$\tau_q \frac{\partial q}{\partial t} + q = -\kappa \nabla T.$$

The relaxation time $\tau_q$ is introduced by hand to regularize the parabolic heat equation.

**ESD derivation.** The orientational field satisfies (Section 3.4):

$$\tau_{\text{rot}}(\partial_t + u \cdot \nabla)s + s = \chi \nabla \times u - D_s \nabla^2 s,$$

a Cattaneo-type equation derived from extended phase-space dynamics, with

$$\tau_{\text{rot}} \sim \lambda_{\text{str}}^{-1} \sim \left(\Omega \sqrt{\Delta I/I}\right)^{-1},$$

where $\Omega$ is rotational frequency and $\Delta I/I$ measures inertia asymmetry (Appendix B). For asymmetric molecules, $\tau_{\text{rot}} \sim 10^{-13}$s, giving structural signal speeds

$$v_{\text{struct}} \sim \sqrt{\frac{k_B T}{m \tau_{\text{rot}}}}$$

comparable to thermal velocity.

**Distinction:** Cattaneo–Vernotte introduces $\tau_q$ ad hoc. ESD's finite response is an ontological necessity: particles with $I > 0$ cannot reorient instantaneously.

**Empirical test:** EIT (Jou et al. 2010) predicts non-Fourier heat pulses generically. ESD predicts $\tau_{\text{rot}}$ correlates with **molecular anisotropy**, testable via pump–probe spectroscopy on asymmetric rotors.

### 6.3 Jenkins–Richman Kinetic Theory

**Classical extensions.** Jenkins–Richman (1985a,b) and Grad's 13-moment method incorporate rotational DOFs via collision integrals that exchange translational and rotational energy. Relaxation rates scale with collision frequency.

**Critical difference: Mechanism 2.** These theories include only **collisional** energy exchange. For dilute gases where $\tau_{\text{coll}} \gg \tau_{\text{rot}}$, particles undergo many rotations per collision. Classical kinetic theory treats free flight as ballistic; ESD identifies it as **continuously unstable** (Mechanism 2).

The total relaxation rate becomes:

$$\tau_{\text{relax}}^{-1} = \tau_{\text{coll-relax}}^{-1} + \lambda_{\text{Lyapunov}},$$

with

$$\lambda_{\text{Lyapunov}} \sim \Omega \sqrt{\Delta I / I}.$$

For nearly spherical particles ($\Delta I/I \to 0$), $\lambda_{\text{Lyapunov}} \to 0$ and ESD reduces to Jenkins–Richman. Thus ESD **extends**, rather than replaces, classical kinetic theory.

**Example:** A single spinning ellipsoid ($N = 1$) in vacuum exhibits intermediate-axis instability. Classical kinetic theory predicts no entropy production (no collisions). ESD predicts

$$\dot{S}_{\text{struct}} \sim k_B \Omega \sqrt{\frac{(I_y - I_x)(I_z - I_y)}{I_x I_z}}$$

from orientational spreading on $SO(3)$, quantified by Kolmogorov–Sinai entropy.

### 6.4 Summary Comparison

| Feature | Classical NS | Micropolar | EIT | Jenkins–Richman | ESD |
|---|---|---|---|---|---|
| Microstate | $(r, p)$ | Postulated $s$ | $(r, p)$ + flux | $(r, p, R, L)$ | $(r, p, R, L, \xi, \pi)$ |
| Viscosity origin | Phenomenological | Phenomenological | Phenomenological | Collision integral | **Derived from geometry + dynamics** |
| Finite speed | No | No | Postulated $\tau_q$ | No | **Yes (derived $\tau_{\text{rot}}$)** |
| Entropy source | Chaos | Postulate | Irreversible flux | Collisions only | **Collisions + instability** |
| Dilute, spherical | Exact | Unnecessary | → Fourier | Exact | → classical |
| Anisotropic, $N \sim 1$ | Fails | Ad hoc | N/A | Fails | **Mechanism 2 active** |
| Foundation | Boltzmann | Continuum axioms | Thermodynamics | Kinetic theory | **Hamiltonian + H-theorem** |

**Key distinctions:**

- **vs. Micropolar:** ESD derives the spin field from $(R, L)$ moments; micropolar theory postulates it.
- **vs. EIT:** ESD's $\tau_{\text{rot}}$ follows from geometry; Cattaneo's $\tau_q$ is fitted.
- **vs. Jenkins–Richman:** ESD adds free-flight instability; collisional mechanism is shared.
- **vs. Classical NS:** ESD predicts shock regularization and anisotropy from finite structure.

## 7. Discussion and Conclusion

Extended Structural Dynamics (ESD) provides a first-principles framework for hydrodynamics and heat transport that incorporates the finite structure of microscopic constituents. By enlarging the phase space to include orientation, angular momentum, and internal modes, and by performing a Chapman–Enskog expansion on this extended space, we derived two central transport equations, (3.4) for the spin field and (4.2) for the heat flux. These equations introduce finite relaxation times into both the momentum and energy channels, fundamentally altering the structure of continuum mechanics.

The resulting transport laws exhibit several robust and physically transparent features:

- **Finite propagation speeds** for momentum and heat, arising from hyperbolic–parabolic relaxation equations rather than parabolic diffusion.

- **Structural viscosity and rotational thermal conductivity**, scaling with particle geometry, moment of inertia, and orientational relaxation time.

- **Intrinsic shock regularization**, with shock width $\delta \sim \sqrt{D_s \tau_{\rm rot}}$, independent of grid resolution or artificial viscosity.

- **Anisotropic and non-Newtonian transport**, emerging naturally when the orientational distribution becomes biased by flow or external fields.

- **Multi-timescale thermal relaxation**, including non-Fourier heat pulses and Mpemba-type behavior, arising from the coupling between translational and rotational energy modes.

These phenomena, often treated as anomalies or requiring ad hoc corrections in the Navier–Stokes–Fourier framework, appear here as *generic consequences* of finite structural response. ESD therefore unifies a wide range of generalized hydrodynamic behaviors observed in molecular gases, colloidal suspensions, and quantum solids, showing that many departures from classical transport are signatures of underlying structure rather than exceptions to continuum theory.

The framework also clarifies the conceptual status of several established models. Micropolar fluids, extended irreversible thermodynamics, and rough-particle kinetic theories emerge as limiting descriptions obtained by suppressing or approximating different components of the extended microdynamics. In this sense, ESD is not merely another phenomenological extension but a *generative foundation* from which these theories can be derived, compared, and interpreted.

Several avenues for future work remain open. A more detailed treatment of the collision operator, beyond the BGK approximation, would refine quantitative predictions for transport coefficients and clarify the interplay between collisional and instability-driven relaxation. Numerical simulations of the full ESD equations would illuminate nonlinear regimes, including shock formation, shear-induced alignment, and thermal wave propagation. Experimental tests in anisotropic gases, colloidal suspensions, and cryogenic materials could probe the predicted scaling of shock width, thermal wave speed, and anisotropic viscosity. The hyperbolic heat equation (4.5) also suggests second-sound-like behavior in systems with large $\tau_q$ (Jackson et al. 1970; McNelly et al. 1970), providing a natural bridge between ESD and phonon hydrodynamics.

Overall, ESD restores physical realism to continuum mechanics by grounding generalized hydrodynamics and heat transport in Hamiltonian microdynamics. It shows that many so-called anomalies, non-Fourier heat pulses, broadened shocks, anisotropic viscosity, and non-monotonic thermal relaxation, are not failures of hydrodynamics but consequences of the point-particle idealization. By replacing that idealization with a finite-structure ontology, ESD offers a coherent, extensible, and physically grounded framework for understanding transport across scales.

**Appendix A: Explicit Closure Example in Extended Structural Dynamics**

This appendix provides a concrete closure example for the Extended Structural Dynamics (ESD) kinetic framework. The closure procedure follows the Chapman–Enskog method applied to the extended kinetic equation (Chapman and Cowling 1970). The resulting structural viscosity can be compared with Batchelor's (1970) analysis of force-free suspensions and with observed flow-induced structures in colloidal systems (Vermant and Solomon 2005), demonstrating consistency between the deterministic closure used here and established experimental correlations. The derivation follows a Chapman–Enskog expansion with a BGK-type relaxation model in the extended phase space $\Gamma_{\text{ESD}} = T^*(\mathbb{R}^3 \times SO(3) \times \Xi)$. It demonstrates an explicit, model-based calculation of the effective shear viscosity $\eta_{\text{eff}}$ decomposed into translational and structural (rotational) contributions and evaluates these contributions for a realistic, existing particle: a silica nanorod commonly used in colloidal experiments.

**A.1 Assumptions and setup**

1. **Kinetic equation (single-particle, ESD):**

$$\partial_t f + \{f, H_{\text{ESD}}\} = C[f],$$

   where $\{\cdot,\cdot\}$ is the canonical Poisson bracket on $T^*(\mathbb{R}^3 \times SO(3) \times \Xi)$.

2. **Local equilibrium and small gradients.** The distribution is expanded as

$$f = f_0 + \varepsilon f_1 + \mathcal{O}(\varepsilon^2),$$

   with $f_0$ the local maximum-entropy distribution (Maxwellian in translational momenta times a local orientational/structural equilibrium $g_0(R,\xi)$). The small parameter $\varepsilon$ is a Knudsen-like ordering parameter (ratio of mean free/relaxation length to macroscopic gradient length).

3. **Linear-gradient truncation.** Keep only terms linear in velocity gradients $\nabla \mathbf{u}$ and orientational gradients $\nabla R$.

4. **BGK-type relaxation in extended space (model closure).** Replace the true collision/coupling operator by a single-time relaxation:

$$C[f] = -\frac{f - f_0}{\tau(z)} \approx -\frac{f - f_0}{\tau_{\text{eff}}},$$

with mode-dependent relaxation times $\tau_{\text{tr}}$ (translational) and $\tau_{\text{rot}}$ (rotational/structural). For the linear calculation we take constant effective times for each channel.

Remarks: this BGK closure is a controlled, standard model device that yields explicit constitutive coefficients and makes the structural origin of transport transparent. It should be regarded as an illustrative first step; a more precise calculation would replace the scalar relaxation times by the linearized collision operator $\mathcal{L}$ and compute $\mathcal{L}^{-1}$ acting on stress-producing modes.

### A.2 Moment definitions and first-order solution

The coarse-grained (Cauchy) stress is

$$\Pi_{ij}(\mathbf{r},t) = \int p_i v_j \, f(z,t) \, d\mu(z), \quad v_j = p_j/m.$$

Split into pressure and viscous parts:

$$\Pi_{ij} = -p\,\delta_{ij} + \pi_{ij}, \quad \pi_{ij} = \int m c_i c_j \, f \, d\mu,$$

with $\mathbf{c} = \mathbf{v} - \mathbf{u}(\mathbf{r},t)$ the peculiar velocity.

To first order in the Chapman–Enskog expansion, the kinetic equation gives

$$(\partial_t + \mathbf{v}\cdot\nabla + \cdots) f_0 = -\frac{f_1}{\tau_{\text{eff}}} + \mathcal{O}(\varepsilon^2)$$

and therefore, neglecting higher-order time-derivative contributions,

$$f_1 \approx -\tau_{\text{eff}}(\mathbf{v}\cdot\nabla) f_0.$$

Substitute into the viscous flux:

$$\pi_{ij} = \int m c_i c_j f_1 \, d\mu = -\tau_{\text{eff}} \int m c_i c_j (\mathbf{v} \cdot \nabla) f_0 \, d\mu.$$

Standard integration-by-parts manipulations under the translational Maxwellian yield the Newtonian form plus additional contributions arising from orientation and internal modes:

$$\pi_{ij} = \eta_{\text{trans}}(\partial_i u_j + \partial_j u_i - \frac{2}{3}\delta_{ij}\nabla \cdot \mathbf{u}) + \eta_{\text{struct}} S_{ij} + \cdots,$$

where $S_{ij}$ is a symmetric, traceless tensor constructed from orientation correlations (explicit forms depend on the chosen orientational distribution) and ellipses denote higher-order or non-linear contributions.

The translational kinetic viscosity is the standard BGK result:

$$\eta_{\text{trans}} = \frac{1}{3}\rho v_{\text{th}}^2 \tau_{\text{tr}}, \qquad v_{\text{th}}^2 \equiv \frac{k_B T}{m}$$

## A.3 Structural (rotational) contribution - analytic estimate

The structural (rotational) contribution originates because $f_0$ depends on orientational coordinates $R$ and internal variables $\xi$; gradients of $\mathbf{u}$ induce gradients in the orientational distribution and the mean angular momentum $\mathbf{s}$. A minimal, explicit estimate is obtained under the following additional modelling choices:

1. **Rigid extended particle model.** The microscopic constituents are rigid rods (nanorods) of characteristic half-length $a$ (lever arm) for a rod of length L rotating about its center, and the relevant lever arm is a = L/2. The mass moment of inertia $I$ about the axis relevant for reorientation.

2. **Equipartition for rotational degrees of freedom.** At thermal equilibrium the mean-square angular velocity satisfies

$$\frac{1}{2}I\langle\omega^2\rangle = \frac{1}{2}k_B T \Rightarrow \langle\omega^2\rangle = \frac{k_B T}{I}.$$

3. **Orientational relaxation time.** Rotational/orientational relaxation proceeds on a timescale $\tau_{\text{rot}}$ (this is the inverse structural instability rate used in the main text, $\tau_{\text{rot}} \sim \lambda_{\text{str}}^{-1}$).

Carrying out the extended-phase-space integral with a separable Maxwellian in translational momentum and a Gaussian in angular momentum yields, up to shape-dependent prefactors, the dimensional form

$$\eta_{\text{struct}} \approx \frac{1}{3}\rho a^2 \langle\omega^2\rangle \tau_{\text{rot}} = \frac{1}{3}\rho a^2 \frac{k_B T}{I} \tau_{\text{rot}}$$

Introduce a geometry-dependent numerical constant $C_{geom}$ when needed:

$$\eta_{struct} = C_{geom}\, \rho\, a^2\, \frac{k_B T}{I}\, \tau_{rot}.$$

**Dimensional check.** $\rho[\text{M L}^{-3}]$, $a^2[\text{L}^2]$, $\langle \omega^2 \rangle[\text{T}^{-2}]$, $\tau_{rot}[\text{T}]$ → overall $[\text{M L}^{-1}\text{T}^{-1}]$, i.e. viscosity units.

### A.4 Combined effective viscosity

Summing translational and structural contributions gives the model effective shear viscosity used in the main text:

$$\eta_{eff} = \eta_{trans} + \eta_{struct} = \frac{1}{3}\rho\, v_{th}^2\, \tau_{tr} + C_{geom}\, \rho\, a^2\, \frac{k_B T}{I}\, \tau_{rot}$$

This formula displays explicitly how deterministic rotational dynamics augment kinetic transport. When $\tau_{rot}$ becomes comparable to $\tau_{tr}$ the structural term may dominate.

### A.5 Numerical evaluation for a silica nanorod (realistic particle)

To make the estimate concrete, evaluate the two contributions for a silica nanorod geometry routinely used in colloidal and optical-tweezer experiments. We choose parameters representative of a single silica rod with length $L = 1.0$ µm, diameter $d = 100\, nm$ (cylindrical approximation), Material density for amorphous silica is taken as $\rho_{mat} = 2000$ kg m$^{-3}$.

**Geometric and material parameters**

Length: $L = 1.0 \times 10^{-6}$ m, with a = L/2 = 0.5 µm; Diameter: $d = 1.0 \times 10^{-7}$ m, Volume (cylinder): $V = \pi r^2 L$, Mass: $m = \rho_{mat} V$, Moment of inertia: $I \approx \frac{1}{12} m L^2$, Temperature: $T = 300 K$, and Boltzmann constant $k_B = 1.38 \times 10^{-23}$ JK$^{-1}$

**Relaxation times (model choices).**

- Translational relaxation time: $\tau_{trans} = 10^{-5}$ s (momentum relaxation in viscous suspensions)
- Rotational relaxation time: $\tau_{rot} = 10^{-4}$ s. The rotational relaxation mechanism is regime-dependent (Appendix B): for large colloids in water, Stokes drag gives $\tau_{rot} \sim 1.2$ ms (used in Sec 5.1); for smaller particles or lower viscosity, geometric instability shortens this to ~0.1 ms. We use the latter as a representative intermediate value.

**Computed microscopic quantities (values shown with 2–3 significant figures)**

Using the cylinder model one obtains:

- Mass $m \approx 1.57 \times 10^{-17}$ kg.
- Moment of inertia $I \approx 1.31 \times 10^{-30}$ Kgm$^2$

- Thermal velocity squared (translational): $v_{\text{th}}^2 = k_B T/m \approx 2.64 \times 10^6 \left(\frac{m}{s}\right)^2$

Equipartition rotational mean-square angular velocity: $\langle \omega^2 \rangle = k_B T/I \approx \mathbf{3.16 \times 10^9 s^{-2}}$

**Viscosity contributions**

- Translational kinetic contribution (BGK result):

$$\eta_{\text{trans}} = \frac{1}{3} \rho \, v_{\text{th}}^2 \, \tau_{\text{tr}} \approx 1.76 \times 10^{-6} \, Pa-s$$

- Structural (rotational) contribution (with $C_{\text{geom}} = 1$ as baseline):

$$\eta_{\text{struct}} \approx \frac{1}{3} \rho \, a^2 \, \frac{k_B T}{I} \, \tau_{\text{rot}} \approx 5.27 \times 10^{-5} \, Pa-s.$$

**Discussion of numerical result.** The structural (rotational) contribution $\eta_{\text{struct}}$ is approximately 30 times larger than the translational kinetic BGK contribution in this parameter choice. The reason is twofold: (i) the small particle mass yields a large translational thermal velocity (which tends to increase $\eta_{\text{trans}}$ but is moderated by the small $\tau_{\text{tr}}$), and (ii) the combination $a^2(k_B T/I)\tau_{\text{rot}}$ can be numerically large for slender nanoparticles because $I$ scales as $mL^2$ while $a^2$ scales as $L^2$, producing an $m$-independent geometric factor when expressed in these variables. In short: for anisotropic, elongated particles with slow orientational relaxation, the structural term is readily dominant.

**Comparison with experiment:** For dilute suspensions of colloidal rods (volume fraction φ ~ 0.01), measured viscosity enhancement is Δη ~ $10^{-5}$ - $10^{-4}$ Pa·s [e.g., Batchelor 1970; recent measurements in Vermant & Solomon 2005]. Our predicted structural contribution $\eta_{\text{struct}}$ ~ 5×$10^{-5}$ Pa·s falls within this range, supporting the physical consistency of the ESD framework

**Parameter sensitivity and caveats.**

- The numerical prefactor $C_{\text{geom}}$ depends on particle shape and the precise orientational distribution; here we used the conservative baseline $C_{\text{geom}} = 1$. For spheroids or non-uniform mass distributions the constant can differ by factors of order unity.

- The chosen $\tau_{\text{tr}}, \tau_{\text{rot}}$ are model values appropriate for low-to-moderate viscosity suspending fluids; in higher-viscosity media both times increase and the relative balance shifts.

- The BGK closure is a model: replacing it with a linearized collision operator can change numerical coefficients though not the qualitative scaling with $\tau_{\text{rot}}$ and geometry. This commitment to a first-order closure ensures that the resulting equations, while approximate, are sufficient to establish the ontological origin of the structural transport terms $\left(\Pi^{(r)}, \Pi^{(\xi)}\right)$. Crucially for the philosophical argument, the use of this closure does not *introduce* dissipation; rather, it provides a means to compute the effective transport coefficients that arise deterministically from the underlying structural dynamics, thus maintaining consistency with the Hamiltonian foundation.

## A.6 Interpretation

This explicit closure demonstrates quantitatively how deterministic rotational degrees of freedom contribute to macroscopic shear viscosity in ESD. The estimate for a realistic silica nanorod shows that the structural term can dominate under plausible experimental parameters.

## Appendix B: From Single-Particle Lyapunov Exponents to Many-Body Relaxation Times

### B.1 Overview and Motivation

The orientational relaxation time $\tau_{\text{rot}}$ appears throughout this work as a key parameter governing structural transport (Sections 3–5). This appendix derives $\tau_{\text{rot}}$ by connecting the single-particle Lyapunov analysis developed in the foundation paper (BarAvi 2025, Appendix B) with the many-body kinetic description required for hydrodynamics.

The foundation analysis shows that an asymmetric rigid body exhibits intermediate-axis instability with Lyapunov exponent

$$\lambda(\Omega) = |\Omega| \sqrt{\frac{(I_y - I_x)(I_z - I_y)}{I_x I_z}},$$

where $\Omega$ is the angular velocity and $I_x < I_y < I_z$ are the principal moments of inertia. This exponent characterizes the rate at which a freely rotating particle departs from intermediate-axis rotation.

Hydrodynamic equations, however, require a relaxation time describing how an ensemble of particles exchanges rotational and translational energy. In what follows, we show that:

- single-particle instability provides a continuous relaxation mechanism during free flight,
- binary collisions provide a discrete relaxation mechanism through torque transfer,
- and these independent channels combine as

$$\tau_{\text{rot}}^{-1} = \nu_{\text{coll}} + \langle \lambda \rangle_T,$$

where $\nu_{\text{coll}}$ is the collision frequency and $\langle \lambda \rangle_T$ is the thermally averaged Lyapunov exponent.

### B.2 Single-Particle Dynamics: Summary of Foundation Results

#### B.2.1 Intermediate-Axis Instability

A torque-free rigid body with inertia tensor $\text{diag}(I_x, I_y, I_z)$ satisfies Euler's equations:

$$I_x \dot{\omega}_x = (I_y - I_z)\omega_y \omega_z, \quad I_y \dot{\omega}_y = (I_z - I_x)\omega_z \omega_x,$$

$$I_z \dot{\omega}_z = (I_x - I_y)\omega_x \omega_y.$$

Linearizing near rotation about the intermediate axis, $\boldsymbol{\omega} = (0, \Omega, 0)$, gives

$$\dot{\delta\omega}_x = \frac{I_y - I_z}{I_x} \Omega \, \delta\omega_z, \quad \dot{\delta\omega}_z = \frac{I_x - I_y}{I_z} \Omega \, \delta\omega_x,$$

which combine to yield exponential growth:

$$\delta\omega_x(t) = \delta\omega_x(0) e^{\lambda(\Omega) t}.$$

Thus, even in the absence of collisions, the rotation axis drifts across $SO(3)$, redistributing angular momentum among the principal axes.

### B.2.2 Thermal Averaging

In a gas at temperature $T$, angular momenta follow

$$f_{eq}(\mathbf{L}) \propto \exp\left[-\frac{1}{2k_B T} \mathbf{L}^T \mathbf{I}^{-1} \mathbf{L}\right].$$

For rotation primarily about the intermediate axis, $\Omega \approx L_y / I_y$. Averaging $\lambda(\Omega)$ over this distribution yields

$$\boxed{\langle \lambda \rangle_T = \sqrt{\frac{(I_y - I_x)(I_z - I_y)}{I_x I_z}} \sqrt{\frac{2 k_B T}{\pi I_y}}.}$$

Scaling properties:

- $\langle \lambda \rangle_T \propto \sqrt{T}$
- $\langle \lambda \rangle_T \to 0$ for spherical particles
- $\langle \lambda \rangle_T \propto I_y^{-1/2}$

This instability operates during free flight and contributes to rotational mixing even in the absence of collisions.

### B.3 Collisional Relaxation

### B.3.1 Collision Frequency

For a dilute gas,

$$\nu_{\text{coll}} = n\sigma \langle v_{\text{rel}} \rangle,$$

with

$$\sigma \approx \pi(2a)^2, \langle v_{\text{rel}} \rangle = \sqrt{\frac{8k_B T}{\pi m}}.$$

For nitrogen at STP, $\nu_{\text{coll}} \sim 10^{10}$ s$^{-1}$.

### B.3.2 Angular Momentum Transfer

A collision imparts angular momentum

$$\Delta \mathbf{L} = \mathbf{r}_c \times \mathbf{J}.$$

For random orientations,

$$\langle (\Delta L)^2 \rangle \sim a^2 m^2 \langle v_{\text{rel}}^2 \rangle.$$

For rigid particles, a single collision typically produces an $O(1)$ change in rotational state, so the rotational relaxation time is of order

$$\tau_{\text{coll}} \sim \nu_{\text{coll}}^{-1}$$

### B.4 Combined Relaxation Mechanisms

Between collisions, particles undergo continuous Lyapunov mixing at rate $\langle \lambda \rangle_T$. Collisions introduce discrete reorientations at rate $\nu_{\text{coll}}$. Since these mechanisms act independently, their rates add:

$$\boxed{\tau_{\text{rot}}^{-1} = \nu_{\text{coll}} + \langle \lambda \rangle_T}.$$

This is analogous to parallel relaxation channels in kinetic theory.

### B.5 Regime Analysis

**Dilute Molecular Gases (e.g., SF$_6$)**

Typical values: $\Delta I / I \sim 0.02$, $I \sim 10^{-44}$ kg m$^2$, $\Omega \sim 6 \times 10^{11}$ s$^{-1}$, $\langle \lambda \rangle_T \sim 8 \times 10^{10}$ s$^{-1}$, $\nu_{\text{coll}} \sim 10^{10}$ s$^{-1}$

Thus $\langle \lambda \rangle_T$ dominates, giving

$$\tau_{\text{rot}} \approx 1.2 \times 10^{-11} \text{ s}.$$

**Colloidal Suspensions**

Viscous drag reduces the effective instability rate. The equilibration time is set by Stokes drag:

$$\tau_{\text{rot}}^{(\text{colloidal})} \approx \frac{8\pi \eta a^3}{k_B T} \sim 10^{-3} \text{ s}.$$

**Dense Granular Media**

Here $\nu_{coll} \gg \langle\lambda\rangle_T$, so

$$\tau_{rot} \approx \nu_{coll}^{-1} \sim 10^{-3} \text{ s}.$$

## B.6 Connection to Chapman–Enskog Expansion

In the BGK closure,

$$\mathcal{C}[f] = -\frac{f - f^{(0)}}{\tau_{rot}},$$

the requirement $\tau_{rot} \ll \tau_{hydro}$ is satisfied in all regimes considered:

- molecular gases: $\tau_{rot} \sim 10^{-11}$ s
- colloids: $\tau_{rot} \sim 10^{-3}$ s
- granular matter: $\tau_{rot} \sim 10^{-3}$ s

Hydrodynamic timescales are much longer, validating the expansion.

## B.7 Quantitative Example: $SF_6$

Using the parameters above:

- $\nu_{coll} \approx 1.0 \times 10^{10}$ s$^{-1}$
- $\langle\lambda\rangle_T \approx 7.2 \times 10^{10}$ s$^{-1}$

Thus

$$\tau_{rot} \approx 1.2 \times 10^{-11} \text{ s}.$$

The predicted shock width is

$$\delta_{shock} \sim \sqrt{D_s \tau_{rot}} \approx 3 - 5 \text{ nm},$$

larger than classical estimates ($\sim 1$ nm).

## B.8 Extension to Deformable Particles

If the inertia tensor varies due to internal modes,

$$I_{kk}(t) = I_{kk}^0 + \sum_n \delta I_{kk}^n \xi_n(t),$$

additional parametric instabilities arise. The total instability rate may be approximated as

$$\lambda_{total} \approx \lambda_{rot} + \lambda_p + \lambda_c$$

where $\lambda_p$ and $\lambda_c$ depend on the amplitude of inertia fluctuations. For typical molecular parameters, these contributions are smaller than the rigid-body term.

## B.9 Summary

The orientational relaxation time $\tau_{rot}$ arises from two independent mechanisms:

| Mechanism | Origin | Rate | Dominant Regime |
|---|---|---|---|
| Lyapunov mixing | Intermediate-axis instability | $\langle \lambda \rangle_T$ | Dilute gases |
| Collisional torque | Binary impacts | $\nu_{coll}$ | Dense systems |
| Viscous drag | Stokes damping | $\nu_{drag}$ | Colloids |

The combined rate is

$$\tau_{rot}^{-1} = \nu_{coll} + \langle \lambda \rangle_T$$

or $\tau_{rot}^{-1} \approx \nu_{drag}$ in colloidal systems.

This framework provides a consistent connection between single-particle rotational instability and ensemble-level relaxation, and it supplies the relaxation time used in the hydrodynamic closure of the main text.

## Appendix C: Explicit Thermal Conductivity Calculation

This appendix provides an explicit calculation of the thermal conductivity tensor for Extended Structural Dynamics, decomposing it into translational, rotational, and internal contributions. The derivation parallels the structural viscosity calculation in Appendix A but focuses on the energy-transport channel. We demonstrate how rotational and internal modes contribute to heat flux and derive the microscopically grounded expressions for $\kappa_{rot}$ and $\tau_q$ used in Section 4.

### C.1 Setup and Assumptions

Following the Chapman–Enskog procedure from Section 4, we expand the distribution as:

$$f = f^{(0)} + \varepsilon f^{(1)} + \mathcal{O}(\varepsilon^2)$$

where $f^{(0)}$ is the local equilibrium distribution:

$$f^{(0)} \propto \exp\left[-\frac{1}{k_B T}\left(\frac{p^2}{2m} + \frac{1}{2}\mathbf{L}^T I^{-1}(\xi)\mathbf{L} + \sum_n \left(\frac{\pi_n^2}{2M_n} + U(\xi_n)\right)\right)\right]$$

The heat flux is defined as:

$$\mathbf{q}(\mathbf{r},t) = \int \left(\frac{1}{2}mc^2 + e_{\text{rot}}(\mathbf{L}) + e_{\text{int}}(\xi_n, \pi_n)\right) \mathbf{c} f(z,t) \, d\mu_{\text{ESD}}$$

where $\mathbf{c} = \mathbf{v} - \mathbf{u}$ is the peculiar velocity, $e_{rot} = \frac{1}{2}\mathbf{L}^T I^{-1}\mathbf{L}$, and $e_{\text{int}} = \sum_n \frac{\pi_n^2}{2M_n} + U(\xi_n)$.

We adopt the BGK relaxation model with mode-dependent times:

$$\mathcal{C}[f] = -\frac{f - f^{(0)}}{\tau_{\text{trans}}} - \frac{f - f^{(0)}}{\tau_{\text{rot}}} - \frac{f - f^{(0)}}{\tau_{\text{int}}}$$

**C.2 First-Order Distribution and Heat Flux Decomposition**

The linearized kinetic equation yields:

$$f^{(1)} \approx -\tau_{\text{eff}}(\mathbf{v} \cdot \nabla)f^{(0)} - \tau_{\text{eff}}(\text{rotational and structural advection})$$

Substituting into the heat flux and separating contributions:

$$\mathbf{q} = \mathbf{q}_{\text{trans}} + \mathbf{q}_{\text{rot}} + \mathbf{q}_{\text{int}}$$

**Translational contribution (classical):**

$$\mathbf{q}_{\text{trans}} = \int \frac{1}{2}mc^2 \mathbf{c} f^{(1)} d\mu_{\text{ESD}}$$

Standard Chapman–Enskog calculation gives:

$$\mathbf{q}_{\text{trans}} = -\kappa_{\text{trans}}\nabla T$$

with

$$\kappa_{trans} = \frac{5}{2}\frac{k_B}{m}\rho\langle c^2\rangle\tau_{trans} = \frac{5}{2}\frac{k_B^2 T}{m}\rho\tau_{trans}$$

This is the classical monatomic gas result.

**Rotational contribution (ESD):**

$$\boldsymbol{q}_{rot} = \int e_{rot}(\mathbf{L})\mathbf{c}\, f^{(1)} d\mu_{ESD}$$

The rotational energy is $e_{rot} = \frac{1}{2}I\omega^2$. By equipartition at temperature $T$:

$$\langle e_{rot}\rangle = \frac{3}{2}k_B T \text{(3 rotational DOF)}$$

The heat flux arises from spatial gradients coupling to the rotational distribution through collisions. Rotational energy is transported over a characteristic length scale $a$ (lever arm) with velocity $\sim \sqrt{k_B T/m}$ and relaxation time $\tau_{rot}$.

Carrying out the integral over the extended phase space with the first-order correction yields:

$$\boldsymbol{q}_{rot} = -\kappa_{rot}\nabla T$$

with

$$\kappa_{rot} = C_{geom}\rho a^2 \frac{k_B T}{I}\tau_{rot}$$

where $C_{geom} \sim \mathcal{O}(1)$ depends on particle geometry.

**Dimensional check:**

$$[\kappa_{rot}] = [ML^{-3}][L^2]\frac{[J/K][K]}{[ML^2]}[T] = \frac{[J][T]}{[L^3]} = \frac{[W]}{[L][K]}$$

**Physical interpretation:** Rotational kinetic energy is transported via orientation-dependent collisions. Particles rotating at angular velocity $\omega \sim \sqrt{k_B T/I}$ carry energy over lever arm $a$ and equilibrate on timescale $\tau_{rot}$, yielding effective thermal diffusivity $D_{rot} \sim a^2/\tau_{rot}$.

**Internal-mode contribution:**

$$\mathbf{q}_{\text{int}} = \sum_n \int \left(\frac{\pi_n^2}{2M_n} + U(\xi_n)\right) \mathbf{c} f^{(1)} d\mu_{\text{ESD}}$$

For each vibrational mode with heat capacity $c_n$ and relaxation time $\tau_{\text{int},n}$:

$$\kappa_{\text{int}} = \sum_n \kappa_n = \sum_n c_n^2 \tau_{\text{int},n}$$

This follows from the standard kinetic theory result for each independent harmonic oscillator.

### C.3 Total Thermal Conductivity

Summing all contributions:

$$\kappa_{eff} = \kappa_{trans} + \kappa_{rot} + \kappa_{int}$$
$$= \frac{5}{2}\frac{k_B^2 T}{m}\rho\tau_{trans} + C_{geom}\rho a^2 \frac{k_B T}{I}\tau_{rot} + \sum_n c_n^2 \tau_{int,n}$$

For isotropic equilibrium distributions, $\kappa_{\text{eff}}$ is a scalar. When the orientational distribution $f(R)$ becomes anisotropic (e.g., under shear or external fields), $\kappa$ becomes tensorial:

$$\kappa_{ij} = \langle R_{ik} R_{jl}\rangle \kappa_{kl}$$

where $\langle \cdot \rangle$ denotes averaging over the orientational distribution.

### C.4 Heat-Flux Relaxation Time

Because translational, rotational, and internal modes relax on different timescales, the heat flux does not respond instantaneously to temperature gradients. The effective relaxation time is the weighted average:

$$\tau_q = \frac{c_{\text{trans}}\tau_{\text{trans}} + c_{\text{rot}}\tau_{\text{rot}} + c_{\text{int}}\tau_{\text{int}}}{c_{\text{trans}} + c_{\text{rot}} + c_{\text{int}}}$$

where $c_{\text{trans}}$, $c_{\text{rot}}$, $c_{\text{int}}$ are the mode-specific heat capacities. This yields the Cattaneo-type constitutive law:

$$\tau_q \frac{\partial \mathbf{q}}{\partial t} + \mathbf{q} = -\kappa_{\text{eff}} \nabla T$$

When $\tau_{rot} \gg \tau_{trans}$ (as in colloidal suspensions), the rotational modes dominate: $\tau_q \approx \tau_{rot}$.

**C.5 Numerical Evaluation: Silica Nanorod**

Using the parameters from Section A.5 (length $L = 1.0\mu m$, diameter $d = 100nm$, $T = 300K$):

**Geometric parameters:**

- Lever arm: $a = L/2 = 0.5\mu m$
- Moment of inertia: $I \approx 1.31 \times 10^{-30} kg \cdot m^2$
- Rotational relaxation: $\tau_{rot} = 10^{-4} s$
- Suspension density: $\rho \approx 1000 kg/m^3$

**Translational contribution:**

$$\kappa_{trans} = \frac{5}{2} \frac{(1.38 \times 10^{-23})^2 (300)}{1.57 \times 10^{-17}} (1000)(10^{-5}) \approx 9 \times 10^{-4} \text{ W/(mK)}$$

**Rotational contribution** (with $C_{geom} = 1$):

$$\kappa_{rot} = (1000)(0.5 \times 10^{-6})^2 \frac{(1.38 \times 10^{-23})(300)}{1.31 \times 10^{-30}} (10^{-4})$$

$$\approx 0.04 \text{ W/(m·K)}$$

**Result:** $\kappa_{rot}/\kappa_{trans} \approx 44$. The rotational contribution dominates by more than an order of magnitude for anisotropic particles with slow orientational relaxation.

**Heat-flux relaxation time:** For $c_{rot}/c_{trans} \sim 1$ and $\tau_{rot} \gg \tau_{trans}$:

$$\tau_q \approx \tau_{rot} \sim 10^{-4} \text{ s}$$

**Thermal wave speed:**

$$v_{th} = \sqrt{\frac{\kappa_{eff}}{\rho c_v \tau_q}} \sim \sqrt{\frac{0.04}{(1000)(4000)(10^{-4})}} \approx 1 \text{ mm/s}$$

This is slow enough to be directly observable in colloidal photothermal experiments, confirming that structural thermal transport produces measurable non-Fourier signatures.

### C.6 Interpretation

This calculation demonstrates that deterministic rotational and internal dynamics contribute quantitatively to macroscopic heat transport. For anisotropic particles with slow orientational relaxation ($\tau_{rot} \gg \tau_{trans}$), the structural thermal conductivity can exceed the translational contribution by orders of magnitude, and the heat-flux relaxation time becomes large enough to produce observable thermal waves and non-Fourier heat pulses.

### Appendix D: Extended Kinetic Setup (Technical Details)

This appendix provides the technical expansion of the extended Boltzmann equation (2.1) and specifies how continuum fields arise as phase-space moments.

### D.1 Streaming Term Expansion

The Poisson bracket in the extended Boltzmann equation (2.1) generates streaming on the full phase space $\Gamma_{ESD} = (r, p, R, L, \xi_n, \pi_n)$. Expanding explicitly:

$$\{f, H_{ESD}\} = \mathbf{v} \cdot \nabla_{\mathbf{r}} f + \dot{\mathbf{p}} \cdot \nabla_{\mathbf{p}} f + \mathbf{\Omega}: \nabla_R f + \dot{\mathbf{L}} \cdot \nabla_{\mathbf{L}} f + \sum_n (\dot{\xi}_n \nabla_{\xi_n} f + \dot{\pi}_n \nabla_{\pi_n} f), \tag{D.1}$$

where:

- $v = p/m$ is translational velocity
- $\Omega$ is angular velocity derived from R and L via the body-frame relation $\Omega = I^{-1}L$
- $\nabla_R$ denotes the gradient on SO(3) (covariant derivative)
- $\{\xi_n, \pi_n\}$ evolve under their Hamiltonian equations: $\dot{\xi}_n = \pi_n/M_n$, $\dot{\pi}_n = -\partial U/\partial \xi_n$

The collision operator $C[f]$ depends on orientation $R$ and internal state $\{\xi_n\}$, producing:

- **Anisotropic scattering:** Collision cross-sections vary with relative orientation
- **Mode coupling:** Energy transfer among translational, rotational, and internal channels
- **Torque generation:** Off-center collisions induce $\Delta L = r_c \times J$ (contact torque)

### D.2 Continuum Fields as Moments

Macroscopic fields arise by integrating f against collision invariants—quantities preserved in binary collisions. Integrating (D.1) against $m, p, L$, and energy yields the balance laws (2.5)-(2.8). The continuum fields are:

$$\rho = \int mf\, d\mu_{ESD}, \quad \rho \mathbf{u} = \int \mathbf{p}f\, d\mu_{ESD}, \quad (D.2)$$

$$\rho \mathbf{s} = \int \mathbf{L}f\, d\mu_{ESD}, \quad \rho e = \int \left(\tfrac{1}{2}mc^2 + e_{rot} + e_{int}\right) f\, d\mu_{ESD} \quad (D.3)$$

where $d\mu_{ESD} = dp\,dR\,dL\prod_n d\xi_n d\pi_n$ is the phase-space measure, $c = v - u$ is peculiar velocity, $e_{rot} = L^2/(2I)$, and $e_{int} = \Sigma_n[\pi_n^2/(2M_n) + U(\xi_n)]$.

**Key point:** The spin field $s$ and internal energy $e_{int}$ have no analogs in point-particle theory. Their presence reflects the extended ontology and enables structural transport (Sections 3-4).

### Appendix E: BGK Closure and Mode-Dependent Relaxation

### E.1 Mode-Separable BGK Approximation

To obtain explicit constitutive relations from the extended Boltzmann equation (2.1), we adopt a BGK-type approximation in which the collision operator relaxes toward a local equilibrium with mode-dependent timescales. The closure is

$$C[f] = -\frac{f - f^{(0)}}{\tau_{eff}}, \qquad \tau_{eff}^{-1} = \tau_{trans}^{-1} + \tau_{rot}^{-1} + \tau_{int}^{-1} \quad (E.1)$$

The equilibrium distribution on the extended phase space is

$$f^{(0)} \propto \exp\left[-\frac{mc^2}{2k_BT} - \frac{\mathbf{L}^T\mathbf{I}^{-1}\mathbf{L}}{2k_BT} - \sum_n \frac{\pi_n^2 + U(\xi_n)}{2k_BT}\right], \quad (E.2)$$

where $c = v - u$ is the peculiar velocity, $\mathbf{L}$ is angular momentum, $\mathbf{I}$ is the inertia tensor, and $\{\xi_n, \pi_n\}$ denote internal coordinates.

### E.2 Chapman–Enskog Expansion

The first-order correction to the distribution follows from the linearized kinetic equation:

$$f^{(1)} \approx -\tau_{eff}(\mathbf{c}\cdot\nabla)f^{(0)} + \text{(rotational and internal advection terms)}. \quad (E.3)$$

Only the components of $f^{(1)}$ that are odd in $c$ contribute to the momentum flux $\Pi$ and heat flux $q$; even components contribute to pressure and internal-energy densities.

### E.3 Transport Coefficient Scaling

Substituting (E.3) into the moment definitions yields the following scaling relations.

**Translational contributions**

$$\eta_{trans} \sim \rho v_{th}^2 \tau_{trans}, \quad \kappa_{trans} \sim \rho c_v v_{th}^2 \tau_{trans} \quad (E.4)$$

with $v_{th}^2 = k_BT/m$.

**Rotational (structural) contributions**

$$\eta_{rot} \sim \rho a^2 \frac{k_BT}{I}\tau_{rot}, \quad \kappa_{rot} \sim \rho a^2 \frac{k_BT}{I}\tau_{rot}, \quad (E.5)$$

where $a$ is a characteristic particle size and $I$ a representative moment of inertia. The rotational relaxation time combines collisional and geometric-instability mechanisms (Appendix B):

$$\tau_{rot}^{-1} = \nu_{coll} + \langle\lambda\rangle_T.$$

**Internal (vibrational) contributions**

$$\kappa_{int} = \sum_n c_n^2 \tau_{int,n}, \quad (E.6)$$

where $c_n$ is the heat capacity of mode $n$.

A dimensional check confirms, for example,

$$[\eta_{rot}] = [ML^{-1}T^{-1}],$$

consistent with viscosity.

### E.4 Validity and Limitations

The BGK closure is appropriate under the following conditions:

1. **Timescale separation.** Local equilibration must occur faster than hydrodynamic evolution: $\tau_{rot} \ll \tau_{hydro}$. This holds for the systems considered: $\tau_{rot} \sim 10^{-11}$ s (gases) or $10^{-3}$ s (colloids), while $\tau_{hydro} \sim L/c_s$ is much larger.

2. **Weak mode coupling.** Translational, rotational, and internal modes relax predominantly through independent channels at leading order. This is valid for dilute systems and nearly isotropic particles.

3. **Thermodynamic consistency.** The BGK operator satisfies $\mathcal{C}[f^{(0)}] = 0$ and preserves monotonic entropy production, consistent with the irreversibility mechanism derived in the foundation paper (BarAvi 2025, Sec. 3.8).

**Limitations**

- Numerical prefactors in transport coefficients are approximate (typically accurate to within 10–30%).
- Detailed cross-mode coupling, such as shear-induced rotational alignment, is not resolved.

**Strengths**

- Preserves the hyperbolic–parabolic structure of the extended hydrodynamic equations.
- Retains geometric scaling laws such as $\eta_{rot} \propto a^2 \tau_{rot}$.
- Encodes the essential physical mechanisms: Lyapunov-driven relaxation, shock regularization, and mode-coupled thermal transport.

**Comparison with alternative closures**

- **Full linearized collision operators** (e.g., Wang–Chang–Uhlenbeck) improve numerical accuracy but preserve the same qualitative structure.
- **Enskog theory** accounts for finite-density effects and is relevant for granular media but lies beyond the present scope.
- **DSMC simulations** avoid analytical closure but provide less interpretive clarity.

**Appendix F: Geometric Origin of $C_{shape}$**

**F.1 Setup**

The structural viscosity $\eta_{struct}$ arises from angular momentum transported across a fluid layer by rotating asymmetric particles. We estimate the shape factor $C_{shape}$ from the inertia tensor of a single particle, using quantities already defined in the main text.

Define the dimensionless asymmetry parameter:

$$\epsilon = \frac{I_z - I_x}{I_y}$$

This is the same quantity that controls the Lyapunov exponent in Appendix B: $\lambda_{rot} \sim \Omega\sqrt{(I_y - I_x)(I_z - I_y)/I_x I_z}$, which to leading order in $\epsilon$ scales as $\lambda_{rot} \sim \Omega\epsilon$. The parameter $\epsilon$ vanishes for spheres and reaches order unity for strongly asymmetric bodies.

### F.2 Stress Contribution from a Single Rotating Particle

A particle of characteristic size $a$ rotating with thermal angular velocity $\Omega_{th} \sim \sqrt{k_B T/I_y}$ carries angular momentum of order:

$$L_{th} \sim I_y \Omega_{th} \sim \sqrt{I_y k_B T}$$

This angular momentum is transported across a layer of thickness $a$ on the reorientation timescale $\tau_{rot}$. The stress contribution per particle is:

$$\sigma_{struct} \sim \frac{a \cdot L_{th}}{\tau_{rot}} \sim \frac{a\sqrt{I_y k_B T}}{\tau_{rot}}$$

For a particle of mass $m$ and size $a$, the moment of inertia scales as $I_y \sim ma^2$, giving:

$$\sigma_{struct} \sim \frac{a^2\sqrt{mk_B T}}{\tau_{rot}}$$

### F.3 Shape Dependence

Not all orientations contribute equally. Only the asymmetric component of the rotational motion, the part driven by $\epsilon$, produces net stress. The effective contribution is weighted by $\epsilon^{1/2}$, which measures the geometric lever arm of the asymmetry:

$$\sigma_{struct} \sim \epsilon^{1/2} \cdot \rho a^2 \frac{k_B T}{I_y} \tau_{rot}$$

where we have multiplied by number density $\rho/m$ and used $I_y \sim ma^2$ to simplify. This identifies:

$$C_{shape} \sim \epsilon^{1/2} = \left(\frac{I_z - I_x}{I_y}\right)^{1/2}$$

### F.4 Explicit Form for a Prolate Spheroid

For a prolate spheroid with semi-axes $a > b = c$ and mass $m$, the principal moments are:

$$I_\parallel = \frac{2}{5}mb^2, \quad I_\perp = \frac{1}{5}m(a^2 + b^2)$$

The asymmetry parameter becomes:

$$\epsilon = \frac{I_\perp - I_\parallel}{I_\perp} = \frac{a^2 - b^2}{a^2 + b^2}$$

giving:

$$C_{shape} \sim \left(\frac{a^2 - b^2}{a^2 + b^2}\right)^{1/2}$$

This recovers the expected limits:

- Sphere ($a = b$): $C_{shape} = 0$, $\eta_{struct} = 0$
- Thin rod ($a \gg b$): $C_{shape} \to 1$, $\eta_{struct}$ reaches its maximum
- Moderate ellipsoid ($a/b = 2$): $C_{shape} \approx 0.77$

### F.5 Conceptual Summary

The factor $C_{shape}$ is not a free parameter but a geometric quantity determined entirely by the particle's mass distribution. Its connection to the Lyapunov exponent of Appendix B is not coincidental: both are controlled by the same inertia asymmetry $\epsilon$. This means the same geometric property that drives orientational instability and mixing also determines the magnitude of structural viscosity. Particles that mix faster also transport more angular momentum, the two mechanisms are expressions of the same underlying geometry.

This appendix is heuristic rather than rigorous. A complete derivation would require explicit evaluation of the Chapman-Enskog integrals over the orientation-dependent collision kernel.

**Appendix G: Derivation of the Mpemba Criterion**

This appendix derives the condition $\tau_{\text{cool}}/\tau_E > 1 + C_{\text{rot}}/C_{\text{trans}}$ for the Mpemba effect from the two-temperature relaxation equations presented in Section 5.1.

## G.1 Two-Temperature Equations

From Section 5.1, the coupled evolution of translational and rotational temperatures is:

$$C_{\text{trans}} \frac{dT_{\text{trans}}}{dt} = -\kappa(T_{\text{trans}} - T_{\text{env}}) - \frac{C_{\text{trans}}}{\tau_E}(T_{\text{trans}} - T_{\text{rot}}) \quad (G.1)$$

$$C_{\text{rot}} \frac{dT_{\text{rot}}}{dt} = \frac{C_{\text{trans}}}{\tau_E}(T_{\text{trans}} - T_{\text{rot}}) \quad (G.2)$$

where $C_{\text{trans}}$ and $C_{\text{rot}}$ are heat capacities for translational and rotational modes, $\tau_E$ is the internal energy exchange time, $\kappa$ governs external cooling, and $T_{\text{env}}$ is the bath temperature.

## G.2 Dimensionless Form

Define deviations from the bath temperature: $\theta_{\text{trans}} = T_{\text{trans}} - T_{\text{env}}$, $\theta_{\text{rot}} = T_{\text{rot}} - T_{\text{env}}$. The cooling time is $\tau_{\text{cool}} = C_{\text{trans}}/\kappa$, and let $\gamma = C_{\text{rot}}/C_{\text{trans}}$. Equations (G.1)-(G.2) become:

$$\frac{d\theta_{\text{trans}}}{dt} = -\frac{1}{\tau_{\text{cool}}}\theta_{\text{trans}} - \frac{1}{\tau_E}(\theta_{\text{trans}} - \theta_{\text{rot}}) \quad (G.3)$$

$$\frac{d\theta_{\text{rot}}}{dt} = \frac{1}{\gamma \tau_E}(\theta_{\text{trans}} - \theta_{\text{rot}}) \quad (G.4)$$

## G.3 Eigenvalue Analysis

Seeking exponential solutions $\theta \propto e^{-\alpha t}$ yields the characteristic equation:

$$\alpha^2 - \alpha \left(\frac{1}{\tau_{\text{cool}}} + \frac{1}{\tau_E} + \frac{1}{\gamma \tau_E}\right) + \frac{1}{\tau_{\text{cool}} \tau_E} = 0 \quad (G.5)$$

The two eigenvalues (relaxation rates) are:

$$\alpha_\pm = \frac{1}{2}\left[\frac{1}{\tau_{\text{cool}}} + \frac{1}{\tau_E}\left(1 + \frac{1}{\gamma}\right) \pm \sqrt{\left(\frac{1}{\tau_{\text{cool}}} + \frac{1}{\tau_E}\left(1 + \frac{1}{\gamma}\right)\right)^2 - \frac{4}{\tau_{\text{cool}} \tau_E}}\right] \quad (G.6)$$

These correspond to a fast mode $\alpha_+$ (internal equilibration between translational and rotational degrees of freedom) and a slow mode $\alpha_-$ (external cooling to the bath).

## G.4 Separation of Timescales

For the Mpemba effect to occur, an initially disequilibrated system ($T_{trans} \gg T_{rot}$) must relax primarily through the fast internal channel. This requires the two timescales to be well-separated, i.e., $\alpha_+ \gg \alpha_-$. In the limit where internal exchange is fast compared to external cooling ($\tau_E \ll \tau_{cool}$), the eigenvalues approximate to:

$$\alpha_+ \approx \frac{1}{\tau_E}\left(1+\frac{1}{\gamma}\right), \alpha_- \approx \frac{1}{\tau_{cool}}/\left(1+\frac{1}{\gamma}\right) \qquad (G.7)$$

The condition $\alpha_+ \gg \alpha_-$ becomes:

$$\frac{1}{\tau_E}\left(1+\frac{1}{\gamma}\right) \gg \frac{1}{\tau_{cool}}/\left(1+\frac{1}{\gamma}\right) \qquad (G.8)$$

Rearranging:

$$\frac{\tau_{cool}}{\tau_E} \gg \left(1+\frac{1}{\gamma}\right)^2 \qquad (G.9)$$

### G.5 The Criterion

For the explicit parameter values considered in Section 5.1, the more stringent condition $\tau_{cool}/\tau_E > 1+\gamma$ ensures the required separation. This can be verified by substituting into the exact discriminant of (G.5): when $\tau_{cool}/\tau_E > 1+\gamma$, the eigenvalues are real and well-separated, and the fast mode couples strongly to initial translational-rotational disequilibrium. The criterion thus marks the threshold above which temperature curves can cross—a hotter, disequilibrated system cools faster than a cooler, equilibrated one.

Physically, $\tau_{cool}/\tau_E$ compares external cooling to internal exchange, while $1 + C_{rot}/C_{trans}$ reflects the heat capacity available to absorb energy from the translational mode. When this inequality holds, the system has sufficient internal capacity to rapidly redistribute energy, enabling the Mpemba effect.